\documentclass[11pt]{article}

\usepackage{graphicx}

\begin{document}

\begin{center}
\textbf{PECULIAR OUTBURST OF THE NEWLY REVEALED VARIABLE STAR V838 
Mon}

\end{center}

\begin{center}
D. Kolev$^{1}$, M. Miko{\l}ajewski$^{2}$, T. Tomov$^{2}$, I. Iliev$^{1}$, J. Osiwa{\l}a$^{2}$,\\ J. Nirski$^{2}$ and C. Ga{\l}an$^{2}$

\end{center}

\begin{center}
$^{1}$ \textit{National Astronomical Observatory Rozhen, Institute of Astronomy, 
BAS, PO Box 136, 4700 Smolyan, Bulgaria,} rozhen@mbox.digsys.bg

$^{2}$ \textit{Centre for Astronomy, Nicolaus Copernicus University, Pl 
87100 Torun, ul. Gagarina 11, Poland,} mamiko@astri.uni.torun.pl

\end{center}
\bigskip 

\noindent
\textbf{ABSTRACT}. {\small We present photometric and high- and low-resolution 
spectroscopic observations of the unusual outburst of V838 Mon. 
The data were collected at the NAO Rozhen, Bulgaria and at the 
Torun Observatory, Poland. Analysis of the peculiar behavior 
of the spectrum of the star is given. The star's radial velocity 
of +60 km/s is derived. A brief discussion concerning the nature 
of the object is also given.}
\medskip

\section{Introduction}

The high-amplitude outburst star activity always focuses the 
attention. It is difficult for detection from the very beginning 
that causes the incomplete knowledge of these phenomena. V838 
Mon is no exception. Brown (2002) first informed about the appearance 
of a new object in Monoceros. Its photographic and visual magnitude 
on January, 6.6-7.6UT was about 10. The first spectra showed 
numerous absorptions features some of them having P Cyg-type 
profile with radial velocities (RV) of $400-500$ km/s (Wagner 2002, 
Della Valle 2002). The object was identified with an anonymous 
star of $V=15.6$ (Munari et al. 2002). The event was puzzle: assumed 
as an extremely slow nova, the star showed no typical spectrum. 
The excess in $(B-V)$ of 0.8 mag corresponded to a distance d\texttt{>}3 
kpc (Zwitter \& Munari 2002). On February, 2 a second, rapid 
outburst begun and on February, 6-7 the maximum of V\ensuremath{\sim}6.5 
was reached. So, the at least ``two-step'' outburst had total amplitude 
of 9 mag! The light curve showed several ``humps'' before a rapid 
decline begun since the mid-April (Fig.1). A light echo, originating 
in circumstellar ring structures was detected (Henden et al. 
2002, Bond et al. 2002, Munari et al. 2002).

\section{Observations }

We undertook spectroscopic and photometric observations within 
the framework of the joint project for investigation of the symbiotic 
stars between Institute of Astronomy, BAS and the Center for 
Astronomy, Torun University. The high-resolution spectra (0.2 
and 0.4 {\AA}) were collected with the coude-spectrograph + LN2-cooled 
CCD camera of the 2m telescope at NAO. The spectral intervals 
were: $4510-4710$ {\AA}{\AA} (1 night, $S/N \sim 30$); $4840-5040$ {\AA}{\AA} (2 nights, $S/N \sim 50$); $5750-5950$ {\AA}{\AA} (4 nights, $S/N \sim 50-100)$ and regions, including H\ensuremath{\alpha} (on all 8 nights, $S/N \sim 80-250)$. One low-resolution (8 {\AA}) slit less grizm-spectrogram was obtained at the Ritchey-Chretien focus using the focal reductor FORERO and the same CCD camera.

Low-resolution spectroscopic observations were carried out in 
the Torun Observatory at Piwnice on 11 nights (see Fig.1). The 
spectra were obtained with the Canadian Copernicus Spectrograph 
+ CCD Peltier-cooling camera, attached to the $F/15$ focus of the 
90 cm Schmidt-Cassegrain reflector. The resolution was 2 {\AA}/pix. 

Both high- and low-resolution frames were processed by standard 
IRAF procedures.

The $UBVR_{c}I_{c}$ photometric observations were carried out using 
the 60cm reflector at Torun Observatory.

\section{Photometric behavior}

Our observations do not cover the whole time interval of the 
event. That is why we use the available published data in order 
to analyze the photometric indicators. The light curve is presented 
in Fig.1. The data set is heterogeneous including standard $UBVRI$, 
unfiltered CCD and simply visual estimates, but nevertheless 
the general shape of the curve is well enough outlined.

The first (at least) rise of the brightness had its maximum near 
the date of revealing the object. The V-brightness rose by 6.5 
mag. Then the star underwent a monotonous decrease from 10 to 
11 mag for 22 days. The reason for the brightening was the expansion 
of stellar pseudo-photosphere as rather slow, optically dense 
stellar wind (indicated by the ``P Cyg"-type profiles 
of the lines). 

It is difficult to estimate the slope of the first outburst. 
If we adopt a brightness of about 13 mag on December 22-25, 2001, 
the linear rate is 0.2 mag/day. Then, for a rise from 15.6 to 
10 mag the star needs 30 days, i.e., the outburst may begin around 
December 10, 2001.

\begin{figure}
\includegraphics[width=\textwidth]{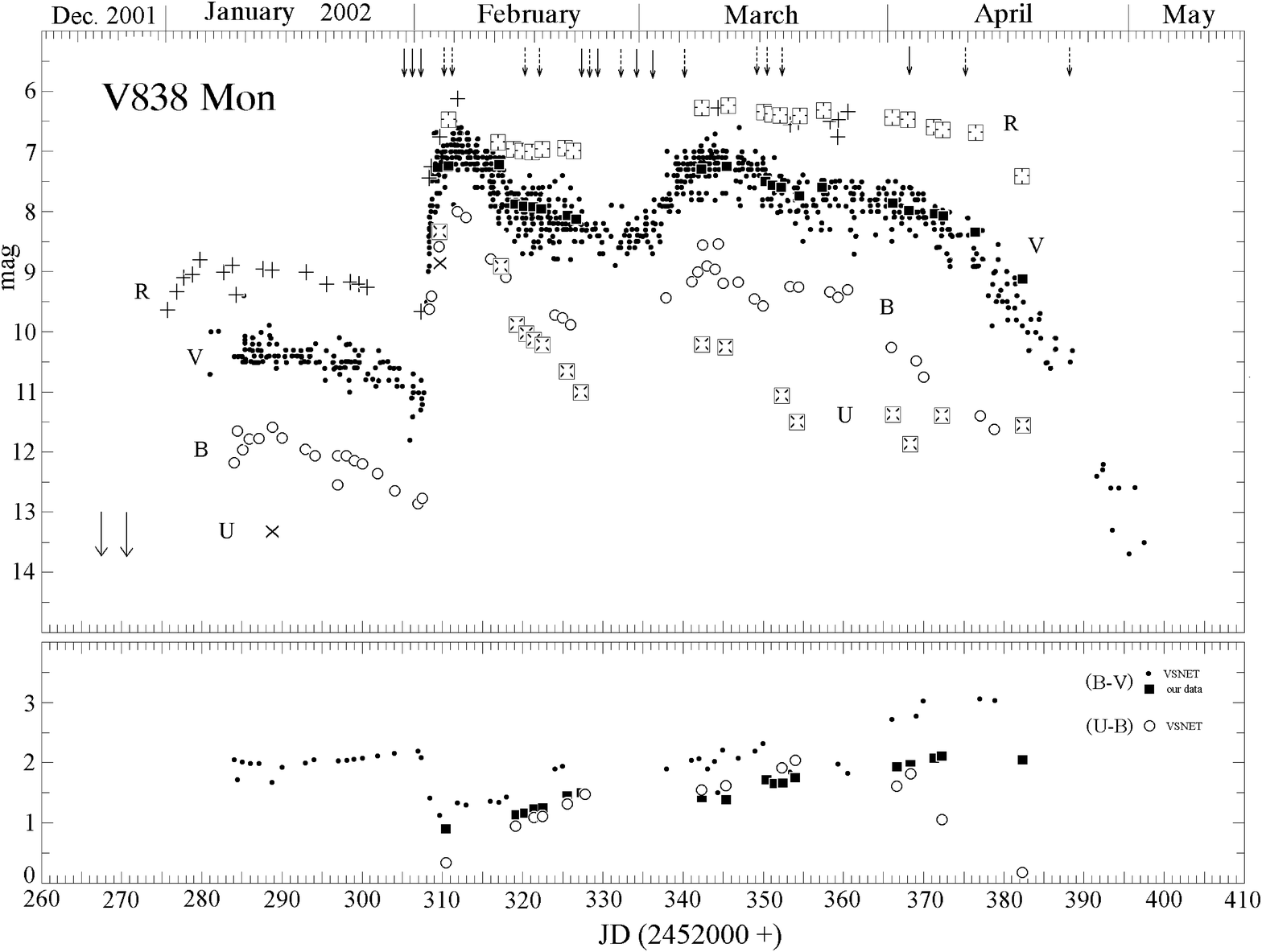}
\caption{Light curve of V838 Mon. The filled 
squares represent our $V-$data; the dots --published in IAU Circulars 
and in the VSNET web site $V-$data; our data in the other colors 
are indicated by squares. In the $B-V$ plot our data are denoted 
by filled squares. The solid and dashed arrows denote the moments 
when high- and low-resolution spectra were obtained.}
\end{figure}

The second, ``main", outburst was very fast and 
the star increased its V-magnitude with amplitude $\Delta V=4$ 
for only 3--4 days. In the same time the color changed its value 
suggesting increase in temperature -- a fact, confirmed by the 
changes in the spectrum (see below). The color reddened again 
when fast drop of the brightness begun on mid-April.

\section{Evolution of the spectrum}

The low-resolution spectra from Jan, 8--10 (Wagner 2002, Della 
Valle 2002) showed that near the maximum of the first outburst 
V838 Mon had a heavily reddened spectrum in which the absorption 
lines of neutral and singly ionized species prevail. Only H\ensuremath{\alpha} 
and few other lines (possibly BaII) had P Cyg profile with weak 
emission components. The measured colors $(B-V)$ corresponded to 
M-giant star.

In the last weak before the second rise of the brightness, the 
spectrum remained reddened (T\ensuremath{\sim}3000K, Lynch et al. 2002) 
and the line spectrum resembled K-giant. Many lines exhibited 
P Cyg shape with deep absorption and relatively weak emission 
components. Narrow H\ensuremath{\alpha}-absorption with \textit{positive} RV 
presented in this time (see Fig. 2). Our spectra show the complex 
shape of all absorption components in P Cyg-profiles of ionized 
metals: the multi-component structure of the absorptions is evident. 
Due to unknown proper velocity of the star and the complexity 
of the RV-systems in the photosphere, the identification of the 
spectrum is very uncertain. There is some possibility to indicate 
line spectra of different excitation temperatures: on our frames 
some high-excitation lines, as NI or HeI can be suspected. Characteristic 
peculiarity of the spectrum of V838 Mon is the presence of numerous ``diffuse interstellar bands" (DIBs). Especially strong and well outlined among them 
are those at \ensuremath{\lambda} 5782 {\AA} and 6379 {\AA} (Fig. 3).

\begin{figure}
\includegraphics[width=\textwidth]{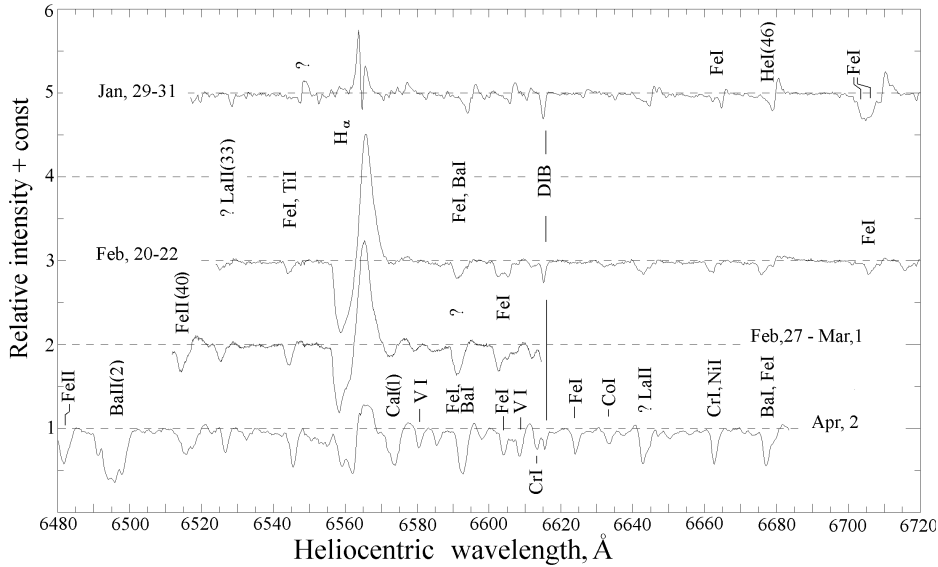}
\caption{High-resolution spectra of V838 Mon near H$\alpha$}
\end{figure}

\begin{figure}
\includegraphics[width=\textwidth]{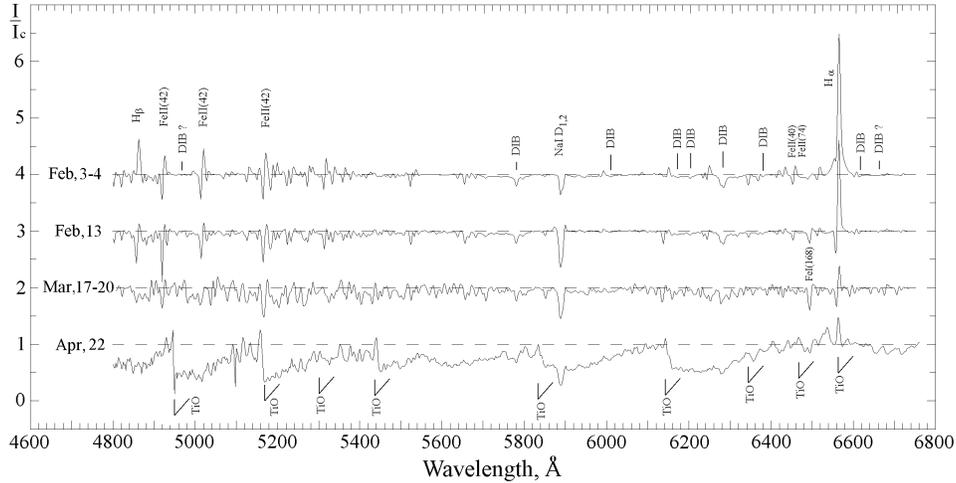}
\caption{Low-resolution spectra of V838 Mon.}
\end{figure}

In the night of the second outburst the spectrum radically changed 
its outlook: the number of ionic lines (FeII, CrII, TiII, SiII, 
etc.) with strong P Cyg-emissions rose significantly. As can 
be seen on our low-resolution spectra (Fig. 3) H\ensuremath{\beta} and 
H\ensuremath{\alpha} showed strong emissions (H\ensuremath{\alpha} exceeded 2.5 
times the continuum level) while one can hardly see the absorption 
component of the lines. The line spectrum resembled rather late 
$A$ or early $F$ star. Two days after the main maximum the color 
$B-V=1.2$ resembled a late $G$ or early $K$ giant or supergiant.

During the next month the spectrum remained at color temperature 
about 4000K (Geballe et al. 2002) while the line spectrum underwent 
serious changes - see Fig.2 and 3: H\ensuremath{\alpha} and many of the 
other lines developed strong P Cyg profiles with deep and relatively 
wide absorptions while the emission components decreased.

Since the mid-March the emission component of H\ensuremath{\alpha} strongly 
decreased (Fig. 3) and even disappeared in H\ensuremath{\beta}. The brightness 
progressively decreased and the photosphere evidently became 
cooler. TiO-bands traces possibly can be found on our spectra 
from March 17--20. 

The spectra obtained on April 9 and 22--23 showed a drastic change 
in the line spectrum (Fig. 3). The M-giant TiO absorption bands 
strengthened, the P Cyg emissions of the near-IR CaII triplet 
disappeared. The only significant emissions remained those of 
H\ensuremath{\alpha}, NaID and several metallic lines. The spectrum looks 
like that of late M giant with a temperature of about 3000K (Henden 
et al. 2002). This temperature drop was interpreted as an expansion 
of the photosphere (Rauch et al. 2002). These spectral changes 
correspond with the drastic drop in the star's brightness showed 
in Fig.1.

\section{Spectral lines from high-resolution spectra}

The evolution of the V838 Mon spectrum demonstrates the unlikeness 
of this event from the known nova-like phenomena. No high-excitation 
emission or forbidden lines appeared; no nebular spectrum was 
developed. Evidently the nature of the star is another and to 
obtain an idea about this phenomenon we must know some star's 
characteristics, such as distance, velocity field in the photosphere, 
etc. The high-resolution frames provide the only data that can 
be of use in such a task.

\subsection{Proper motion}

The complex shape of the spectrum of V838 Mon makes identification 
of all features extremely difficult. The absence of symmetric, 
pure absorption lines in our spectra does not allow measuring 
the proper radial velocity of the star. A possible solution can 
give the Gaussian-fitting of the H\ensuremath{\alpha}-profile obtained 
on the end of January (see Fig. 2). The reason for such an attempt 
gives the fact, that only in the spectra from this period more 
or less symmetric emission component, \textit{not flanked by absorption}, 
appeared. The emission ($FWHM=2.4$ {\AA}, $I \sim 2I_{cont}$) has a 
heliocentric \ensuremath{\lambda}=6564.155 {\AA} and narrower ($FWHM=1$ {\AA}), 
deep ($R_{c}=-1.1$) absorption is centered on 6564.447 {\AA}. Assuming 
spherical emitting volume one can propose the radial velocity 
shift of the emission $RV_{e}=+60$ km/s to reflect the real proper 
radial velocity of the star $RV_{s}$. The radial velocity of the 
absorption is $RV_{a}=+75$ km/s. The close $RV$ of both H\ensuremath{\alpha} 
emission and absorption features possibly argue the presence 
of absorbing stationary or rotating zone around the star.

\subsection{Interstellar (and circumstellar?) NaID lines}

The heavily reddened continuum observed, the numerous strong 
DIBs and the pronounced interstellar (IS) components of NaID 
lines (see Fig. 4) witnesses at first glance for great distance 
to the star. The discovery of the light echo indicating the presence 
of surrounding nebula or cocoon additionally complicates the 
situation. In the same time this event fortunately allowed to 
estimate the distance in an independent from the interstellar 
absorption way. Henden et al. (2002) evaluated the distance d=700 
pc and finally Munari et al. (2002) fixed the distance to 790\ensuremath{\pm}30 pc.

The analysis of the IS-lines components also can give useful 
information concerning the absorption toward the star. There 
exists a well established relation between the mean equivalent 
width $EW$ (in {\AA}) of both D1(5896 {\AA}) and D2(5890 {\AA}) lines 
and the distance (in kpc): $d=2 \times EW$ (Allen 1973). The region around 
NaID lines from our high-resolution spectra is presented in Fig. 
4. In fact each of both IS D-lines has a complex shape, clearly 
resolved in 0.1 {\AA}/pix spectra obtained on Feb, 27 and Mar, 1 
(the solid line in Fig.4). The significantly wider ($FWHM=0.6$ 
{\AA}) blue-edge component surely indicate even more complex structure 
of the interstellar matter toward V838 Mon. Both IS-components \textit{a, 
b} of each D-line have mean $RV$ of +37.2 km/s and +63.9 km/s. The 
$EW$ of the components for these three periods of observation lie 
between $0.81-0.89$ {\AA} for D1 and $0.90-1.00$ {\AA} for D2. The mean 
$EW$ are 0.84 and 0.94 {\AA} respectively with mean $EW=0.89$ {\AA}. According 
to the above cited relation this value correspond to a distance 
$d=1.8$ kpc if a ``normal" interstellar reddening 
law is adopted. The discrepancy of this value from the established 
by the light-echo is evident. Even more significant is the difference 
from the early-established d\texttt{>}3 kpc.

\begin{figure}
\includegraphics[width=\textwidth]{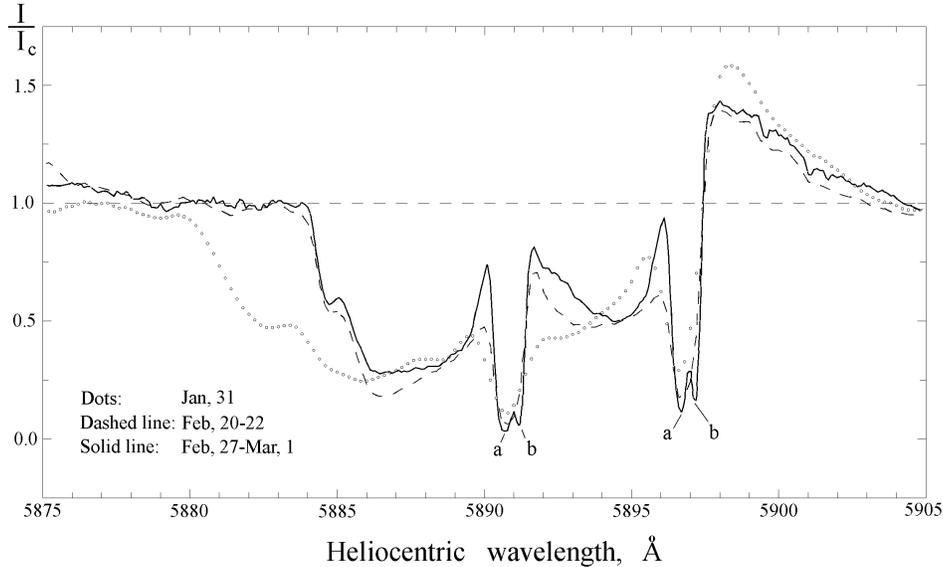}
\caption{NaID lines in the spectrum of V838 Mon.}
\end{figure}

Munari et al. (2002) briefly discussed the interstellar reddening 
toward V838 Mon. They concluded that the color excess for this 
star lies between 0.25 (if $d=0.79$ kpc) and $\sim 0.80$ ($d=3$ 
kpc). The adopted value $E(B-V)=0.5$ is confirmed as well by their 
polarimetric observations. We agree that the presence of \textit{circumstellar} 
mater changes in unpredictable way the absorption and the reddening 
may differ significantly from the ``standard" one. 
The circumstellar gas and dust that contributes to the strength 
of the IS-lines or of the DIBs must have some radial velocity, 
characterizing the mass-outflow. If the proposition for mass-loss 
from the star with a velocity of 15 km/s (Zwitter \& Munari 2002) 
is true, there must be a component with $\Delta RV=-15$ km/s 
according to the $RV_{s}$. Having $RV_{s}=+60$ km/s one can expect the 
presence of a component with $RV \sim +45$ km/s. Our measured 
$RVs$ coincide with the velocities measured by Zwitter \& Munari 
(2002) after the heliocentric correction of about $-10$ km/s for 
their date of observation -- January, 26. \textit{So we can conclude, 
that the wider IS-component ($RV=+37$ km/s) can be affected} \textit{or 
even} \textit{can be fully generated} \textit{by the circumstellar outflow 
material}. If we exclude this component from the estimations, 
the mean $EW$ of the narrower component of 0.33 {\AA} gives d\ensuremath{\sim}0.7 
kpc! Despite this formal closeness to the light-echo derived 
distance, we must reject the NaID interstellar lines as an indicator 
of the distance. Nevertheless the analysis of the shape of the 
Na-lines allows us to conclude, that the contribution from the 
rich in matter environment of the star certainly can be responded 
also for the unusually strong DIBs that may be simply diffuse \textit{circumstellar} 
bands.

\subsection{Outflow velocities}

The spectral lines showed most complex shape just before the 
main outburst. The mean of six FeII lines (well seen in Fig. 
2) radial velocity profile in our end-of-January frames can be 
fitted with at least 4 absorption components. They have heliocentric 
$RV$ of $-15$, $-70$, $-125$ and $-175$ km/sec (with boundary near $-250$ 
km/sec). The emission component has $RV$ about $+50-55$ km/sec. The 
$FWHM$ ($100-115$ km/sec) of this emission coincides with that of 
the H\ensuremath{\alpha}-emission discussed above in {\S}5.1.

The ``typical" $RVs$ of the metallic lines in the end of February 
were between $-80 - -100$ km/sec while the most rapid components 
showed $RV$ of $-270$ or even $-450$ km/sec (very week features). The 
hydrogen lines absorptions were with $RV$ of $-170$ (H\ensuremath{\beta}) 
and $-220$ (H\ensuremath{\alpha}) km/sec. The same value for H\ensuremath{\alpha} 
was registered also on our April 2 frame.

\section{Discussion: what is V838 Mon?}

We already noted the strange behavior of V838 Mon. Different 
authors tried to find analogues to the event: from ``\textit{peculiar 
slow nova or post-AGB star in outburst}'' (Della Valle \& Iijima 
2002), ``\textit{a close analogue of FG Sge, V605 Aql and V1334 Sgr}" 
(Henden et al. 2002a) to an analogue of ``\textit{the very luminous 
extra galactic red variable `M31RV' that had a long outburst 
in 1988}" (Bond et al. 2002). Comparing the light curve 
shape of V838 Mon with other known eruptive objects, we can note 
the following. The summary amplitude of the event, $\Delta V=9$ 
mag and the shape of the curve are similar to that of some known 
Novae. The important nova's parameter $t_{3}$ (the interval since 
the moment of maximum until the decreasing of brightness by 3 
mag) for V838 Mon is $\sim 72$ days. Similar value for $t_{3}$ 
(70 days) and for the amplitude ($16.0-6.5$ mag) had the fast nova 
HR Lyr (1919) (Allen 1973). The comparison with FG Sge shows 
drastic discrepancies: the brightness of FG Sge increased almost 
linearly from 14 to 9.5 mag (amplitude of only 4.5 mag) for more 
than 60 years (Herbig \& Boyarchuk 1968)! The V838 Mon' light 
curve resembles also the curve of the eruptive variable HM Sge 
(Ciatti et al. 1977). The last star was classified as early PN-stage 
or as an object related to V1016 Cyg and V1329 Cyg, but not as 
a typical symbiotic star (Arkhipova et al. 1979), while now it 
is considered to be a symbiotic nova (Schmid 2000). The symbiotic 
nova RR Tel has amplitude of 10 mag. A number of other kinds 
of variable luminous stars show comparable amplitudes. So, Mira 
Cet changes its magnitude from 2 to 10 mag. But the spectral 
behavior that V838 Mon demonstrated differs completely from the 
spectral evolution of classical and symbiotic novae or Miras. 

There can be also found more ``contra'' than ``pro'' arguments comparing 
V838 Mon with the other cited objects. Possibly, the suggestion 
proposed by Munari et al. (2002) that we are looking at new class 
objects may be a clue to the understanding of V838 Mon?

\noindent
\textbf{Acknowledgments}. We are grateful to Drs. D. Kyurkchieva and 
D. Marchev from the Shumen University for obtaining a spectrum 
on April 2 and Drs. T. Bonev and E. Semkov from IA BAS for low 
resolution FORERO-frame. This study was partly supported by Polish 
KBN Grants 5 P03D 003 20 and 5 P03D 013 21.
\bigskip 

\noindent \begin{large}\textbf{References}\end{large}
\medskip 

\noindent
Allen C.W., 1973, Astrophysical quantities, 3$^{rd}$ ed., {\S}108, Univ. of London,

The Athlone Press

\noindent
Arkhipova V.P., Dokuchaeva O. D., Esipov V.F., 1979, Astron Zh, Vol.56, 

\#2, 313

\noindent
Bond H.E., Panagia N., Sparks W.B., Starrfield S.G. and Wagner R.M., 

2002, "HST observations of V838 Mon light echo" (in vsnet-campaign-

v838mon 363 from May, 4)

\noindent
Brown N. J., 2002, IAU Circ. 7785

\noindent
Ciatti F., Mammano A., Vittone A., 1977, A\&A, 61, 459

\noindent
Della Valle M., Iijima T., 2002, IAU Circ. 7786

\noindent
Geballe T.R., Evans A., Smalley B., et al., 2002, IAUCirc. 7855

\noindent
Henden A., Munari U., Schwartz M., 2002a, IAU Circ. 7859

\noindent
Henden A., Munari U., Marrese P., et al., 2002b, IAU Circ. 7889

\noindent
Herbig G.H., Boyarchuk A., 1968, ApJ, 153, 397

\noindent
Lynch D.K., Rudy R.J., Mazuk S., et al., 2002, IAU Circ. 7829

\noindent
Munari U, Henden A, Kiyota S., et al., 2002, A\&A, 389, L51

\noindent
Rauch T., Kerber F., van Wyk F., et al., 2002, IAUCirc. 7996

\noindent
Schmid H.M., 2000, "Observations of hot stellar winds in symbiotic sys-

tems", in: Thermal and Ionization Aspects of Flows from Hot Stars: 

Observations and Theory. ASP Conference Series, Vol. 204, eds. 

H. J.G.L.M. Lamers and A. Sapar, p. 303

\noindent
Wagner R. M., 2002, IAU Circ. 7785

\noindent
Zwitter T., Munari U., 2002, IAU Circ. 7812

\end{document}